\documentclass[aps,pra,floatfix,superscriptaddress,showpacs,amssymb]{revtex4}
\usepackage{amsmath}
\usepackage{mathrsfs}
\usepackage{graphicx}
\usepackage{dcolumn}
\usepackage{bm}
\usepackage{euscript}

\def\prn#1{{\left(#1\right)}}
\def\cbrk#1{{\left\{#1\right\}}}
\def\sbrk#1{{\left[#1\right]}}

\def\abs#1{{\left|#1\right|}}

\def\eqnref#1{{Eq. \eqref{#1}}}

\def\ts#1{_{\mbox{\scriptsize #1}}}
\def\tu#1{^{\mbox{\scriptsize #1}}}

\def\ket#1{{|#1\rangle}}
\def\bra#1{{\langle#1|}}

\def\cg(#1,#2)(#3,#4)(#5,#6){\bra{#1,#2,#3,#4}#5,#6\rangle}
\def\threej(#1,#2)(#3,#4)(#5,#6){\begin{pmatrix}#1&#3&#5\\#2&#4&#6\end{pmatrix}}
\def\sixj(#1,#2,#3)(#4,#5,#6){\begin{Bmatrix}#1&#2&#3\\#4&#5&#6\end{Bmatrix}}
\def\ninej(#1,#2,#3)(#4,#5,#6)(#7,#8,#9){\begin{Bmatrix}#1&#2&#3\\#4&#5&#6\\#7&#8&#9\end{Bmatrix}}

\def\sR{{\ensuremath{\EuScript R}}}

\usepackage{color}

\begin{document}

\pagenumbering{arabic}

\setcounter{footnote}{0}

\title{Hyperfine-interaction- and magnetic-field-induced Bose-Einstein-statistics suppressed two-photon transitions}

\author{M.~G.~Kozlov}
\affiliation{Petersburg Nuclear Physics
Institute, Gatchina, 188300, Russia}

\author{D.~English}
\affiliation{Department of Physics, University of California at
Berkeley, Berkeley, California 94720-7300}

\author{D. Budker}
 \email{budker@berkeley.edu}
 \affiliation{Department of Physics, University of California at
Berkeley, Berkeley, California 94720-7300}
 \affiliation{Nuclear Science Division, Lawrence
 Berkeley National Laboratory, Berkeley, California 94720}

\date{\today}

\begin{abstract}
Two-photon transitions between atomic states of total electronic angular momentum $J_a=0$ and $J_b=1$ are forbidden when the photons are of the same energy. This selection rule is analogous to the Landau-Yang theorem in particle physics that forbids decays of vector particle into two photons. It arises because it is impossible to construct a total angular momentum $J_{2\gamma}=1$ quantum-mechanical state of two photons that is permutation symmetric, as required by Bose-Einstein statistics. In atoms with non-zero nuclear spin, the selection rule can be violated due to hyperfine interactions. Two distinct mechanisms responsible for the hyperfine-induced two-photon transitions are identified, and the hyperfine structure of the induced transitions is evaluated. The selection rule is also relaxed, even for zero-nuclear-spin atoms, by application of an external magnetic field. Once again, there are two similar mechanisms at play: Zeeman splitting of the intermediate-state sublevels, and off-diagonal mixing of states with different total electronic angular momentum in the final state. The present theoretical treatment is relevant to the ongoing experimental search for a possible Bose-Einstein-statistics violation using two-photon transitions in barium, where the hyperfine-induced transitions have been recently observed, and the magnetic-field-induced transitions are being considered both as a possible systematic effect, and as a way to calibrate the measurement.
\end{abstract}


\maketitle

\section{Introduction}

Among the selection rules for two-photon transitions \cite{Bon84,Dun04,DeM00}, there is a peculiar rule that forbids to all orders $J_a=0\rightarrow J_b=1$ transitions when the two photons are collinear and degenerate (i.e., when their frequencies are the same), even when the transition is allowed for non-degenerate photons. This selection rule has the same origin as the Landau-Yang theorem \cite{Lan48,Yan50} that forbids a vector particle, i.e., a particle with intrinsic angular momentum one, to decay into two photons. It arises because, for two photons, it is impossible to construct a quantum-mechanical state that would correspond to total angular momentum $J_{2\gamma}=1$ and would be symmetric with respect to permutation of the two photons, as required by Bose-Einstein (B-E) statistics.

It is just this selection rule that is the basis of the experiment \cite{DeM99,DeM00} with two-photon atomic transitions in barium that has searched for and set stringent limits on a possible small violation of the B-E statistics for photons. A more recent version of the experiment \cite{Bro00,Eng07,Eng09}, using an improved experimental technique, has further tightened the limit on a possible statistics violation for photons. The probability for two 556-nm photons to be in a ``wrong'' permutation-symmetry state has been constrained to be less than $3\cdot10^{-11}$.

In the experiment of Refs.\ \cite{Eng07,Eng09},
two independent tunable narrow-band cw dye lasers with orthogonal polarizations are locked to an in-vacuum optical power-buildup cavity (PBC).  An atomic beam of barium, moving perpendicularly to the PBC optical axis, passes through the coincident waists of the laser beams at the cavity's center.  The sum of the frequencies of the photons from the two lasers is scanned over the frequency of the two-photon resonance between the ground $6s^2\ ^1S_0$ and the excited $5d6d\ J_b=1$ states (relevant energy levels of Ba I are listed in Tab.\ \ref{Tab1}). Fluorescence from the upper state is monitored.

When the two lasers are frequency locked to the same optical mode of the PBC, the photons are degenerate, and the transition is forbidden by the above-mentioned selection rule. However, when the lasers are locked to different modes of the cavity, there arises a non-zero transition probability that scales as the square of the frequency difference between the two lasers. This signal is used to calibrate the sensitivity of the experiment to a possible forbidden transition.

Of the seven stable barium isotopes, only two have non-zero nuclear spin $I$: $^{135}$Ba (6.6\% natural abundance, $I=3/2$) and $^{137}$Ba (11.2\%, $I=3/2$). The present paper is a theoretical investigation of how hyperfine interactions relax the strict suppression of degenerate two-photon transitions. We identify two distinct mechanisms that are responsible for the hyperfine-interaction-induced two-photon transitions: hyperfine splitting of the intermediate state of the transition, and off-diagonal mixing of the states of different total electronic angular-momentum (mostly state $b$ in the case of Ba).

We note that, while we are not aware of any previous studies of hyperfine-interaction-induced (HFI) two-photon transitions, there are several other situations where hyperfine interactions render non-zero amplitudes to forbidden transitions. These include $J$-forbidden transitions relevant to atomic clocks based on trapped ions and neutral atoms, forbidden transitions in highly-charged ions, and highly-suppressed magnetic-dipole transitions of relevance to atomic parity-violation experiments (see, for example, Problems 1.11 and 3.18 in Ref.\ \cite{BudBook2008} and references therein).

Apart from hyperfine interactions, the degenerate two-photon-transition selection rule is also relaxed, even for zero-nuclear-spin atoms, by application of an external magnetic field. This is related to the modification of the Landau-Yang theorem in the presence of a magnetic field considered in Ref.\ \cite{Ang2007}. Once again, there are two mechanisms at play here as in the case of the HFI transitions: Zeeman splitting of the intermediate-state sublevels, and off-diagonal mixing of the final state with states of different total electronic angular momenta. The former effect has been investigated in Ref.\ \cite{Auz2007} using numerical methods. Below, we present an analytical treatment of both effects.

\section{HFI transitions: a qualitative discussion}
\begin{figure}
\begin{center}
\includegraphics[scale=1]{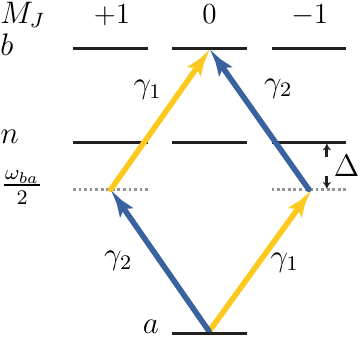}
\end{center}
\caption{(color online). A two-photon transition between $J_a=0$ and $J_b=1$ states must proceed via a virtual intermediate state with $J_n=1$. There are two quantum paths between the initial and the final state that differ by the order of absorption of the photons, and which cancel each other in the case of degenerate photons. The transition to one particular upper-state Zeeman component ($M_J=0$) is shown as an example.}\label{Fig1}
\end{figure}

The B-E suppression of the $J_a=0\rightarrow J_b=1$ degenerate two-photon transitions can be understood as destructive interference of two alternative quantum paths connecting the initial and the final state, as illustrated in Fig.\ \ref{Fig1}. While degenerate two-photon transitions between the states of total angular momentum ($F_a\rightarrow F_b$) other than $0\rightarrow 1$ are not B-E-statistics forbidden; in the absence of hyperfine mixing and energy shifts, the presence of the nonzero-spin nucleus does not allow the transition because the underlying electronic transition is still the degenerate $J_a=0\rightarrow J_b=1$ transition.

\begin{figure}
\begin{center}
\includegraphics[scale=1]{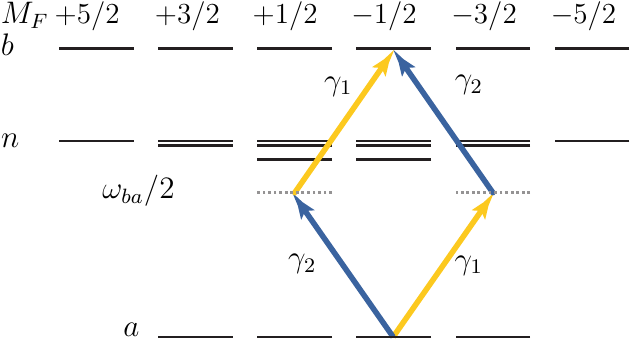}
\end{center}
\caption{(color online). With the addition of nuclear spin $I=3/2$, hyperfine-structure splitting in the intermediate states $n$ may lift the cancelation between the two quantum paths of the two-photon transition. Zeeman splitting of the intermediate state has a similar effect (see text).}\label{Fig2}
\end{figure}

\begin{figure}
\begin{center}
\includegraphics[scale=1]{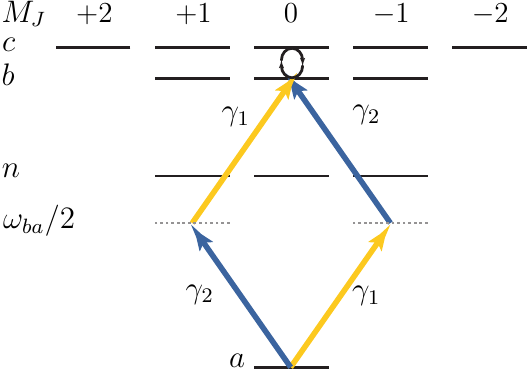}
\end{center}
\caption{(color online). Off-diagonal hyperfine interaction mixes the upper state with nearby states with $J_c\ne 1$, which leads to a non-vanishing degenerate two-photon amplitude. Upper-state mixing can also be induced by an external magnetic field.}\label{Fig3}
\end{figure}

We identify two distinct mechanisms by which this destructive interference can be perturbed. Consider a transition between specific hyperfine levels. As illustrated in Fig.\ \ref{Fig2}, the transition proceeds via intermediate states which, due to hyperfine splitting of the intermediate $J_n=1$ state, have slightly different energy defects with respect to the energy of a photon. This results in slightly different energy denominators associated with the quantum paths that would otherwise cancel. This constitutes the first mechanism via which the hyperfine interactions induce the two-photon transition. The second mechanism (illustrated in Fig.\ \ref{Fig3}) is off-diagonal mixing of states of different total electronic angular momenta. While, in principle, both the initial and the final states can be mixed, the effect in the transitions of interest in Ba is dominated by the mixing of the final state $b$ with nearby states $c$ of the same-parity with $J_c=0$ and 2.

\section{Calculation of the HFI amplitudes}
\subsection{The general expression}
The starting point of our calculation is the general expressions for the amplitude of a degenerate two-photon (E1-E1) transition (see the derivation of  similar expressions in Ref.\ \cite{Dem98}) between specific hyperfine-structure components $F_a$ and $F_b$ of the initial and the final level:
\begin{align}
W_{b,a}=\sum_{\kappa=0}^2\sum_{Q=-\kappa}^{\kappa}(-1)^Q A^{\kappa}_Q P^{\kappa}_{-Q}.
\end{align}
Here $\kappa$ is the tensorial rank, whose range (0-2) is determined by the possible values of the total angular momentum associated with a system of two photons;
\begin{align}
P^{\kappa}_{-Q}=\sum_{q_1,q_2=-1}^{1}\cg(1,q_1)(1,q_2)(\kappa,-Q)\epsilon^1_{q_1}\epsilon^2_{q_2},
\end{align}
where $\cg(1,q_1)(1,q_2)(\kappa,-Q)$ are the Clebsch-Gordan coefficients, is a tensor built out of the polarization vectors of the two light fields ($\epsilon^1_{q_1},\epsilon^2_{q_2}$ are the spherical components of the polarization vectors), and where
\begin{align}
\label{Eq:AkappaQ}
A^{\kappa}_Q=(-1)^{F_b-M_b}\threej(F_b,-M_b)(\kappa,Q)(F_a,M_a)A^{\kappa};
\end{align}
\begin{align}
 A^{\kappa}&=(-1)^{2I+2F_a+F_b}\sqrt{(2\kappa+1)(2F_a+1)(2F_b+1)}
 \sum_{F_n=|J_n-I|}^{J_n+I}(-1)^{2J_n+F_n}(2F_n+1)\times \notag\\
 &\sixj(F_n,F_b,1)(\kappa,1,F_a)\sixj(J_n,F_n,I)(F_b,J_b,1)
 \sixj(J_n,F_n,I)(F_a,J_a,1)
 ||d_{an}||\times||d_{bn}||
 \prn{\frac{1+(-1)^{\kappa}}{\frac{E_b+E_a}{2}-E_n}}.
\label{Eq:Akappa}
\end{align}
In this expression, $I$ is the nuclear spin; a single intermediate state of total electronic angular momentum $J_n$ is assumed (otherwise, a summation over intermediate states should be done); $\cbrk{}$ denote $6j$ symbols; $||d||$ denote the reduced electric-dipole matrix elements in the $J$ basis and do not depend on the total angular momentum $F$.  Expression \eqref{Eq:Akappa} can be derived in a straightforward way using angular-momentum theory (Ref.\ \cite{Sob92}, for example).

\subsection{Intermediate-state-splitting effect}
Examining the expression in the parentheses of Eq.\ \eqref{Eq:Akappa}, we see that the amplitudes of odd
rank $\kappa$ identically vanish for the present case of degenerate two-photon
transitions. In the absence of the hyperfine interactions, all other
amplitudes also vanish for the case of
$J_a=0,\,J_b=1$. For zero spin $I=0$, we have $F_a=0,\,F_b=1$ and the triangle
rules for Eq.\ \eqref{Eq:AkappaQ} require $\kappa=1$. For nonzero nuclear spin, the sum
over $F_n$ in Eq.\ \eqref{Eq:Akappa} turns to zero for $\kappa\neq 1$ even
though individual contributions may be finite.

However, this is no longer the case when we take into account the hyperfine splitting in the
intermediate state:
\begin{align}
E_n=\frac{E_b+E_a}{2}+\Delta + \frac{A_n}{2}C_n + \frac{B_n}{8}\prn{\frac{3C_n(C_n+1)-4I(I+1)J(J+1)}{I(2I-1)J(2J-1)}}, \label{Eq:HFSenergies}
\end{align}
where $\Delta$ is the energy difference between the state $n$ (before including the HF splitting) %
and the mid-point between the energies of the states $a$ and $b$.  The last two terms in \eqnref{Eq:HFSenergies} are the HF splitting; $A_n$ and $B_n$ are the magnetic-dipole and electric-quadrupole hyperfine constants, and
\begin{align}
C_n=F_n(F_n+1) - J_n(J_n+1) - I(I+1).
\end{align}
Because of energy shifts of Eq.\ \eqref{Eq:HFSenergies}, the energy denominators in Eq.\ \eqref{Eq:Akappa}
now depend on $F_n$, and the sum over $F_n$ is not necessarily zero. The amplitudes of the specific $F_a\rightarrow F_b$ components of the
transition $J_a=0\rightarrow J_b=1$ are first order in $A_n/\Delta$ and/or
$B_n/\Delta$.

\begin{table}
\begin{center}
\begin{tabular}{lllclr@{.}lr@{.}lr}
\hline Label & Parity & $E$, cm$^{-1}$ & $E-E_b$, cm$^{-1}$ & Designation & \multicolumn{2}{c}{$A$, MHz }& \multicolumn{2}{c}{$B$, MHz }& Ref.\\
\hline\hline
$a$ & Even & 0          &      & $6s^2\ ^1S_0$ &\multicolumn{2}{l}{} &\multicolumn{2}{l}{} &                \\
$n$ & Odd  & 18 060.261 &      & $6s6p\ ^1P_1$ & -98&16(14)          &  34&01(22)          & \cite{van95}   \\
$b$ & Even & 35 933.806 &      & $5d6d\ ^3D_1$ & -103&7(6)           &  -6&9(7)            & \cite{Eng09}   \\
$c$ & Even & 35 616.949 & -317 & $6s7d\ ^3D_2$ &  \multicolumn{1}{r}{298}&(5)           &  14&7(5)            & \cite{Jit79}   \\
$c'$& Even & 35 762.187 & -172 & $6s7d\ ^1D_2$ &   34&1(3)           &   \multicolumn{1}{r}{3}&(2)            & \cite{Jit79}   \\
    & Even & 36 200.412 &  267 & $5d6d\ ^3D_2$ &\multicolumn{1}{r}{22}&(1)           & \multicolumn{1}{r}{-10}&(3)          & \cite{Jit79}   \\
 \hline
\end{tabular}%
\end{center}
\caption{Relevant energy levels in Ba I \cite{Cur2004} and the hyperfine-structure constants for $^{135}$Ba.}
\label{Tab1}
\end{table}

\subsection{Off-diagonal mixing effect}
A small hyperfine-interaction-induced admixture to the upper state $F_b$ of a
state of the same parity and \emph{total} angular momentum, but with a total
\emph{electronic} angular momentum
$J_c\neq J_b$,
can be described by a mixing
coefficient
\begin{align}
\sbrk{\sixj(F_b,I,J_c)(1,J_b,I)A_{cb} + \sixj(F_b,I,J_c)(2,J_b,I)B_{cb}}\frac{(-1)^{J_b+I+F_b}}{E_b-E_c},\label{Eq:MixCoeff}
\end{align}
where $A_{cb}$ and $B_{cb}$ are the off-diagonal magnetic-dipole and
electric-quadrupole hyperfine mixing coefficients, respectively. This form follows directly from perturbation theory and angular-momentum algebra.
The two-photon transition amplitude induced by the mixing effect is calculated as the product of this
mixing coefficient and the $A^{\kappa}$ of Eq.\ \eqref{Eq:Akappa} with a
substitution of
$J_b\rightarrow J_c$ and $||d_{bn}||\rightarrow||d_{cn}||$. The resulting amplitudes are first order
in $A_{cb}/(E_b-E_c)$ and/or $B_{cb}/(E_b-E_c)$.

For the Ba transition of present interest, there are three states (see Tab. \ref{Tab1})
with $J=2$ close to the final state $5d6d\ ^3D_1$, namely $5d6d\
^3D_2$, $6s7d\ ^1D_2$. and $6s7d\ ^3D_2$. The energy separations for
all three states are comparable (fourth column of Tab.\ \ref{Tab1}). Note that here we follow level assignments from Ref.\ \cite{Cur2004}. In Ref.\ \cite{Kul01}, the level at 35762 cm$^{-1}$ is listed as $^3D_2$, not $^1D_2$. As we derive below, a large magnetic hyperfine splitting is expected for a triplet state of the $6s7d$ configuration, so the fact that the splitting is small supports the term assignment of Ref.\ \cite{Cur2004}. Note also that various properties of these and other nearby levels, including measurement of their unusually high electric polarizabilities, polarization-dependent photoionization cross-sections, and lifetimes have recently been reported in Refs.\ \cite{Li2004,Li2006}.

Hyperfine interactions are sensitive to the wave function near the origin. The $6d$ and $7d$ orbitals are rather weakly bound, defuse orbitals.
Because of this, their contribution to the hyperfine amplitudes is strongly suppressed, so that the main contribution to the hyperfine amplitudes between states of interest should come from the $6s$ and $5d$ orbitals. The former does not contribute to the quadrupole term but gives the largest contribution to the magnetic term. The latter contributes to both terms, but these contributions are suppressed by the strong centrifugal barrier. Below we assume that the magnetic $6s$ amplitude is much larger than both $5d$ amplitudes.

Hyperfine interactions are short-range, so they are the strongest for the lowest allowed partial wave. For the magnetic interaction, the largest contribution comes from the $s$-wave. The electric-quadrupole interaction for $s$-electrons is zero, so the dominant contribution, in this case, comes from the $p$-wave. In the single-particle approximation, going to the next partial wave typically results the loss of the strength of the interaction by an order of magnitude. Because of this, hyperfine constants for $d$-states are usually dominated by electron-correlation effects (see, for example, Refs.\ \cite{Dzu98,Por99}). In the approximation where only the $s$ electrons contribute, and where we ignore the contributions that require configuration mixing in both states (i.e., the contributions proportional to two small mixing amplitudes), we can neglect the mixing between the final state $b$ ($5d6d\ ^3D_1$) and the state $5d6d\ ^3D_2$ and focus on the admixtures of states $c$ ($6s7d\ ^3D_2$) and $c'$ ($6s7d\ ^1D_2$).

Let us start with the hyperfine mixing with the $c$ state. Nominally, the levels $b$ and $c$ can not be mixed by hyperfine interactions, which are described by a one-electron operator that can only mix configurations that differ by one electron at most. However, the HFI mixing is allowed by configuration mixing.

According to the configuration-mixing analysis of Ref.\ \cite{Kul01}, the two relevant states can be written as
\begin{align}
\ket{b} &= \sqrt{0.73} \ket{5d6d\ ^3D_1} +  \sqrt{0.064} \ket{6s7d\ ^3D_1},\\
\ket{c} &= \ket{6s7d\ ^3D_2},
\end{align}
so that the state $\ket{c}$ can be considered pure.

The magnetic-dipole hyperfine-interaction operator can generally be written as $\hat{H}\ts{hfs}=\vec{I}\cdot\vec{V}$, where $\vec{V}$ is a pseudo-vector operator related to the electronic spin and orbital angular-momentum operators \cite{Sob92}. With this, we write (using formulae given in Ch.\ 4 of Ref.\ \cite{Sob92}):
\begin{align}
\bra{c}\hat{H}\ts{hfs}\ket{b}&=\sqrt{0.064}\bra{6s7d\ ^3D_2} \hat{H}\ts{hfs} \ket{6s7d\ ^3D_1}\nonumber\\
&=\sqrt{0.064} (-1)^{1+I+F_b}\sixj(F_b,I,2)(1,1,I)\bra{6s7d\ ^3D_2}|V|\ket{6s7d\ ^3D_1}\bra{I}|I|\ket{I}.\label{Eq:OffDiagME}
\end{align}
The nuclear reduced matrix element is $\bra{I}|I|\ket{I}=\sqrt{I(I+1)(2I+1)}$, and
we can relate the reduced matrix element of $\vec{V}$ to the hyperfine-structure constant of the level $\ket{c}$ by writing a formula for the hyperfine shift of the hyperfine component of the state $\ket{c}$ with total angular momentum $F_b$ in a way analogous to Eq.\ \eqref{Eq:OffDiagME}:
\begin{align}
\bra{c}\hat{H}\ts{hfs}\ket{c}&=(-1)^{2+I+F_b}\sixj(F_b,I,2)(1,2,I)\bra{6s7d\ ^3D_2}|V|\ket{6s7d\ ^3D_2}\bra{I}|I|\ket{I}.\label{Eq:DiagME1}
\end{align}
On the other hand, from the definition of the hyperfine constant $A_c$, we also have
\begin{align}
\bra{c}\hat{H}\ts{hfs}\ket{c}&=A_c\bra{c}\vec{I}\cdot\vec{J}\ket{c} = A_c \frac{F_b(F_b+1)-J_c(J_c+1)-I(I+1)}{2}.\label{Eq:DiagME2}
\end{align}
Comparing Eqs.\ \eqref{Eq:DiagME1} and \eqref{Eq:DiagME2}, for example, for a specific case of $F_b=3/2$, we obtain
\begin{align}
\bra{6s7d\ ^3D_2}|V|\ket{6s7d\ ^3D_2}=\sqrt{30}A_c.\label{Eq:||V||}
\end{align}
Finally, we need to relate the diagonal and off-diagonal reduced matrix elements of $\vec{V}$ in Eqs.\ \eqref{Eq:OffDiagME} and \eqref{Eq:||V||}.
From the Wigner-Eckart theorem, we can write the matrix elements for specific $M_{J_c}=1$ components:
\begin{align}
\bra{^3D_2,1}\vec{V}\ket{^3D_2,1}=(-1)^{2-1}\threej(2,-1)(1,0)(2,1)\bra{^3D_2}|V|\ket{^3D_2},\\
\bra{^3D_2,1}\vec{V}\ket{^3D_1,1}=(-1)^{2-1}\threej(2,-1)(1,0)(1,1)\bra{^3D_2}|V|\ket{^3D_1},
\end{align}
where, for compactness, we are no longer explicitly writing the electron configurations.
Using these equations and Eq.\ \eqref{Eq:||V||}, we obtain:
\begin{align}
\bra{^3D_2}|V|\ket{^3D_1}=\frac{\bra{^3D_2,1}\vec{V}\ket{^3D_1,1}}{\bra{^3D_2,1}\vec{V}\ket{^3D_2,1}}\sqrt{10}A_c.\label{Eq:RedVandA}
\end{align}
Explicitly, the  $6s7d\ ^3D_{J_c},M_{J_c}=1$ states can be written as
\begin{align}
\ket{^3D_2,1}&=\frac{1}{\sqrt{3}}\ket{s\downarrow}\ket{d_2\downarrow}+\frac{1}{2\sqrt{3}}\ket{s\uparrow}\ket{d_1\downarrow}+
\frac{1}{2\sqrt{3}}\ket{s\downarrow}\ket{d_1\uparrow}-\frac{1}{\sqrt{2}}\ket{s\uparrow}\ket{d_0\uparrow},\\
\ket{^3D_1,1}&=\sqrt{\frac{3}{5}}\ket{s\downarrow}\ket{d_2\downarrow}-\frac{\sqrt{3}}{2\sqrt{5}}\ket{s\uparrow}\ket{d_1\downarrow}-
\frac{\sqrt{3}}{2\sqrt{5}}\ket{s\downarrow}\ket{d_1\uparrow}+\frac{1}{\sqrt{10}}\ket{s\uparrow}\ket{d_0\uparrow},\label{Eq:3D11}
\end{align}
where we used the notation in which the projection of each of the electrons is designated by an up or a down arrow. We can now explicitly evaluate the matrix elements
\begin{align}
\bra{^3D_2,1}\vec{V}\ket{^3D_2,1}&=\prn{\frac{1}{3}+\frac{1}{12}}\bra{s\downarrow}\vec{V}\ket{s\downarrow}
+\prn{\frac{1}{2}+\frac{1}{12}}\bra{s\uparrow}\vec{V}\ket{s\uparrow}\nonumber\\
&=\frac{1}{6}\bra{s\uparrow}\vec{V}\ket{s\uparrow};\label{Eq:3D2V3D2}\\
\bra{^3D_2,1}\vec{V}\ket{^3D_1,1}&=\prn{\frac{1}{\sqrt{5}}-\frac{1}{4\sqrt{5}}}\bra{s\downarrow}\vec{V}\ket{s\downarrow}
+\prn{-\frac{1}{4\sqrt{5}}-\frac{1}{2\sqrt{5}}}\bra{s\uparrow}\vec{V}\ket{s\uparrow}\nonumber\\
&=-\frac{3}{2\sqrt{5}}\bra{s\uparrow}\vec{V}\ket{s\uparrow}.\label{Eq:3D2V3D1}
\end{align}
Substituting these results into Eq.\ \eqref{Eq:RedVandA}, we obtain
\begin{align}
\bra{^3D_2}|V|\ket{^3D_1}=-9\sqrt{2}A_c,
\end{align}
and, substituting into Eq.\ \eqref{Eq:OffDiagME}, we get
\begin{align}
\bra{c}\hat{H}\ts{hfs}\ket{b}\approx 3.2(-1)^{I+F_b}\sixj(F_b,I,2)(1,1,I)\sqrt{I(I+1)(2I+1)}\times A_c.\label{Eq:Final}
\end{align}
Comparing Eqs.\ \eqref{Eq:Final} and \eqref{Eq:MixCoeff}, we obtain:
\begin{align}
 A_{cb} \approx -3.2\sqrt{I(I+1)(2I+1)}\times A_c.\label{Eq:OffDiag3D2}
\end{align}

Calculation of the mixing of the state $b$ with the state $c'$ ($6s7d\ ^1D_2$) can be done in the same way. If we write the
wave function for $M_{J_{c'}}=1$ as
\begin{align}
\ket{^1D_2,1}&=\frac{1}{\sqrt{2}}\prn{\ket{s\uparrow}\ket{d_1\downarrow}
-\ket{s\downarrow}\ket{d_1\uparrow}},\label{Eq:1D21}
\end{align}
we find that, under the adopted approximations, the hyperfine splitting of the singlet state $c'$ vanishes as
\begin{align}
\bra{^1D_2,1}\vec{V}\ket{^1D_2,1}&=\frac{1}{2}\sbrk{\bra{s\uparrow}\vec{V}\ket{s\uparrow}+\bra{s\downarrow}\vec{V}\ket{s\downarrow}}=0.
\end{align}
Experimentally, we indeed find that the magnetic hyperfine constant for the state $c'$ is an order of magnitude smaller than those for the nearby triplet states of the same configuration. Next, using Eqs.\ \eqref{Eq:3D11} and \eqref{Eq:1D21} we evaluate the off-diagonal matrix element
\begin{align}
\bra{^1D_2,1}\vec{V}\ket{^3D_1,1}=-\frac{\sqrt{3}}{2\sqrt{10}}\sbrk{\bra{s\uparrow}\vec{V}\ket{s\uparrow}-\bra{s\downarrow}\vec{V}\ket{s\downarrow}}
=-\sqrt{\frac{3}{10}}\bra{s\uparrow}\vec{V}\ket{s\uparrow}.\label{Eq:1D2V3D1}
\end{align}
From Eqs.\ \eqref{Eq:1D2V3D1} and \eqref{Eq:3D2V3D2}, we have:
\begin{align}
\frac{\bra{^1D_2,1}\vec{V}\ket{^3D_1,1}}{\bra{^3D_2,1}\vec{V}\ket{^3D_2,1}}=-\frac{6\sqrt{3}}{\sqrt{10}}.\label{Eq:1D2V3D1over3D2V3D2}
\end{align}
Finally, using the Wigner-Eckart theorem, and taking into account Eq.\ \eqref{Eq:||V||}, we obtain for the reduced matrix elements of $\vec{V}$
\begin{align}\label{Eq:1D2||V||3D1}
\bra{^1D_2}|V|\ket{^3D_1}=-6\sqrt{3}A_c,
\end{align}
and from an expression analogous to Eq.\ \eqref{Eq:OffDiagME} combined with Eq.\ \eqref{Eq:MixCoeff},
\begin{align}
 A_{c'b} \approx -2.6\sqrt{I(I+1)(2I+1)}\times A_c.\label{Eq:OffDiag1D2}
\end{align}
Since the energy intervals between the state $b$ and the states $c$ and $c'$ are comparable, Eqs.\ \eqref{Eq:OffDiag3D2} and \eqref{Eq:OffDiag1D2} indicate that the states $c$ and $c'$ are mixed into the state $b$ in comparable amounts, despite the smallness of the hyperfine splitting in the singlet state $c'$.

\section{Results for specific light polarizations}
Using the formulas derived above, we now perform specific calculations for the two-photon transition in $^{135}$Ba and $^{137}$Ba as an example. We envision an experimental arrangement where two counter-propagating laser beams interact with barium atoms. We assume that during the transition, a single photon is absorbed from each of the beams. While it is possible for two photons from the same beam to be absorbed, these two scenarios can be distinguished by their different spectral profiles: Doppler-free in the former case, and Doppler-broadened in the latter.

In order to develop intuition for the relative importance of various effects, we first calculate the transition rates in terms of the magnitude-squares of the irreducible amplitudes $A^\kappa$ (see Table \ref{Tab:GeomFac}) summed over all possible final magnetic sublevels and averaged over the initial sublevels.  Then the values of the amplitudes for the splitting and mixing effects are calculated (see Table \ref{Tab:Amps}) using the known values of the hyperfine-structure constants for the single, dominant, intermediate state $6s6p\ ^1P_1$ at $18 060.261\ $cm$^{-1}$ (see Table \ref{Tab1}).  In Table \ref{Tab:Amps}, we have multiplied the calculated values of $A^\kappa$ by $\Delta^2(\Delta/A\ts{norm})^2$, where $\Delta=(E_b+E_a)/2-E_n$, and $A\ts{norm}$=100 MHz, so that ``1'' in the resulting units roughly corresponds to a two-photon transition probability suppressed by $(\Delta/A_n)^2\approx 10^{9}$ compared to an allowed two-photon transition with similar parameters such as, for example, the separation $\Delta$.

\def\arraystretch{1.4}%
%
%
\begin{table}
\begin{center}
\begin{tabular}{|r@{\,}l|l|l|l|}%
  \hline
  $\hat{\epsilon}_1$ & $\hat{\epsilon}_2$ & $F_b$ & Rate \\\hline
$\hat x$&$\hat z$& $\frac{1}{2},\frac{3}{2},\frac{5}{2}$ & $\frac{1}{120} \left(5 \abs{A^1}^2+3 \abs{A^2}^2\right)$ \\
$\hat z$&$\hat z$& $\frac{1}{2},\frac{5}{2}$ & $\frac{1}{120}\, 4\abs{A^2}^2$ \\
$\hat z$&$\hat z$& $\frac{3}{2}$ & $\frac{1}{120} \prn{10\abs{A^0}^2+4\abs{A^2}^2}$ \\
$\hat{\sigma}_+$&$\hat{\sigma}_-$ & $\frac{1}{2},\frac{5}{2}$ & $\frac{1}{120} \left(5 \abs{A^1}^2+\abs{A^2}^2\right)$ \\
$\hat{\sigma}_+$&$\hat{\sigma}_-$ & $\frac{3}{2}$ & $\frac{1}{120} \left(10 \abs{A^0}^2+5 \abs{A^1}^2+\abs{A^2}^2\right)$ \\
$\hat{\sigma}_+$&$\hat{\sigma}_+$ & $\frac{1}{2},\frac{3}{2},\frac{5}{2}$ & $\frac{1}{120}\,6 \abs{A^2}^2$ \\
  \hline
\end{tabular}
\end{center}
\caption{Resonant degenerate two-photon transition rate per atom for $F_a=3/2\:\to\:F_b$, where $\hat{\epsilon}_1$ and $\hat{\epsilon}_2$ are the photon polarizations, and $A^\kappa$ is the rank--$\kappa$ irreducible component of the total amplitude (Eq.\ \eqref{Eq:Akappa}).  The expressions are general, for any amplitudes that are irreducible of rank 0-2.  In our application, $A^1=0$.  The values of $A^\kappa$, for $\kappa\neq1$, are found in Table \ref{Tab:Amps}.}
\label{Tab:GeomFac}
\end{table}

%
%
\begin{table}
\begin{center}
\begin{tabular}{|l|l|D{.}{.}{2}|D{.}{.}{2}@{\,}c@{}D{.}{.}{1}@{\,}l|D{.}{.}{2}|D{.}{.}{2}@{\,}c@{}D{.}{.}{1}@{\,}l|}%
  \hline
  \multicolumn{2}{|l|}{}&\multicolumn{5}{c|}{$^{135}$Ba}&\multicolumn{5}{c|}{$^{137}$Ba}\\
  \hline
$F_b$ & $\kappa$ &
\multicolumn{1}{c|}{$A^{\kappa}\ts{Split}$} & \multicolumn{4}{c|}{$A^{\kappa}\ts{Mix}$}  &
\multicolumn{1}{c|}{$A^{\kappa}\ts{Split}$} & \multicolumn{4}{c|}{$A^{\kappa}\ts{Mix}$}  \\\hline
$\frac{3}{2}$ & $0$ &  2.1  &  0 &\multicolumn{3}{l|}{}      &  2.3 &  0 &\multicolumn{3}{l|}{}\\
$\frac{1}{2}$ & $2$ & -0.77 &-0.72&$\sR_c$\,&-\,1.1&$\sR_{c'}$& -1.0 & -0.81&$\sR_c$\,&-\,1.2&$\sR_{c'}$ \\
$\frac{3}{2}$ & $2$ &  1.6 & 1.8&$\sR_c$\,&+\,2.7&$\sR_{c'}$&  2.0 &  2.0&$\sR_c$\,&+\,3.1&$\sR_{c'}$ \\
$\frac{5}{2}$ & $2$ & -1.5 &-2.6&$\sR_c$\,&-\,3.9&$\sR_{c'}$& -1.6 & -2.9&$\sR_c$\,&-\,4.3&$\sR_{c'}$ \\
  \hline
\end{tabular}
%
%
\end{center}
\caption{Resonant degenerate two--photon transition amplitudes due to splitting ($A^{\kappa}\ts{Split}$) of the intermediate states and mixing ($A^{\kappa}\ts{Mix}$) in the final states, calculated in $^{135}$Ba and $^{137}$Ba.  The ratios $\sR_c=||d_{nc}||/||d_{nb}||$, $\sR_{c'}=||d_{nc'}||/||d_{nb}||$, have not been measured.  All amplitudes have been multiplied by ${\Delta ^2}/{A\ts{norm} ||d_{an}||\,||d_{nb}||}$, where $\Delta=(E_b+E_a)/2-E_n$, and $A\ts{norm}$=100 MHz.  The total amplitude $A^{\kappa}$ is the sum of $A^{\kappa}\ts{Split}$ and $A^{\kappa}\ts{Mix}$, but the relative sign between them is unknown.  Untabulated amplitudes are zero.
}
\label{Tab:Amps}
\end{table}

\begin{figure}
\begin{center}
\includegraphics[scale=1]{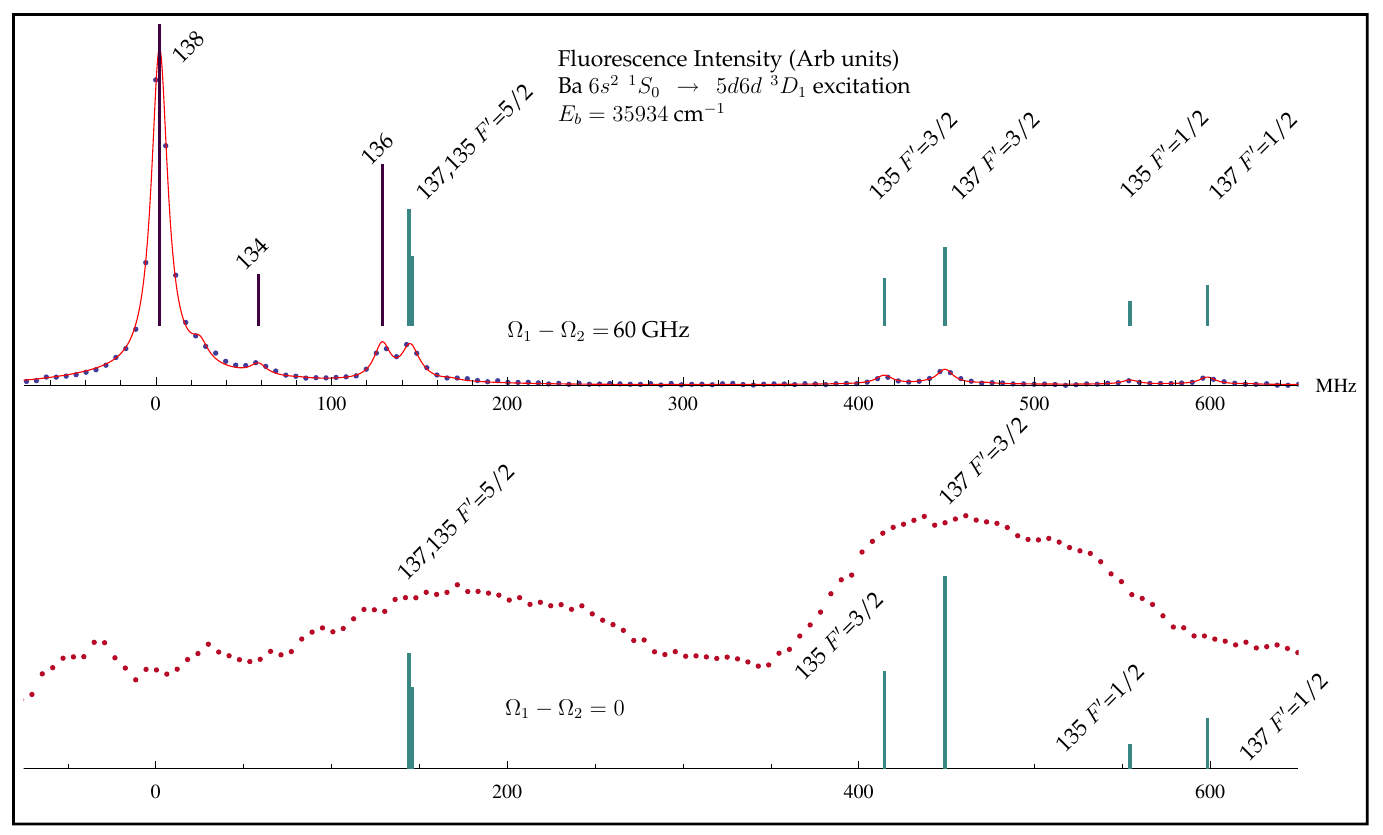}
\end{center}
\caption{(color online). Fluorescence intensity of Ba $5d6d\:^3D_1$ during two-photon excitation.  Upper trace shows non-degenerate excitation: The two photons are separated in frequency by 60 GHz.  Lower trace shows degenerate excitation:  The two photons have the same frequency. The horizontal frequency variable is $\Omega_1+\Omega_2-E_b$ of $^{138}$Ba .  ``Sticks'' indicate the line-positions of the isotopic and hyperfine components.  Stick heights indicate the calculated relative intensities of the peaks. The two traces are plotted on the same frequency scale (with a possible mismatch of no more than 50 MHz).  The vertical scale of the lower plot is expanded $\sim 100\times$ the upper.  The excitation intensity is $\sim10^3$ to $10^4$ times higher in the case of the lower spectrum than in the upper one.  The broad asymmetric line shapes in the lower trace are due to the AC Stark effect (see Sec. \ref{APBCDoesWhat}).  Comparison of upper and lower traces shows that the nuclear-spin zero isotopes disappear during degenerate excitation, and that the relative intensities of the hyperfine components change, in qualitative agreement with calculations.}\label{Fig:modelfit}
\end{figure}

\subsection{What happens in a power-buildup cavity?}\label{APBCDoesWhat}
In the experiments of Refs.\ \cite{Eng07,Eng09}, light with orthogonal linear polarizations from two single-frequency cw dye lasers is coupled, from opposite directions, into a Fabry-Perot power-buildup cavity. Assuming equal intensities and frequencies of the two light beams and an ideal high-finesse cavity, each of the input beams establishes a standing wave in the cavity with a corresponding linear polarization. A superposition of two such waves with orthogonal polarizations is also a standing wave; however, the polarization of the resultant wave depends on the (arbitrary) phase between the two laser beams, and could be any elliptical polarization with a restriction that a principal axis of the polarization ellipse be at $\pi/4$ to each of the laser polarizations. Thus, the experimental situation can be described as when the two photons in the above formulas are of \emph{the same} elliptical polarization. The time-averaged signal can be found by averaging the calculated signal over the relative phase of the two laser fields.


Because the two-photon transition rate goes as the product of the two lasers' intensities, a power-buildup cavity greatly increases (by a
factor of $\sim10^4$ in the case of \cite{Eng07,Eng09}) the excitation rate when the transition is allowed.  But the enhancement comes at
the expense of a complicated lineshape:  Atoms enter the laser beams with a distribution of velocities and entry points, and sample different intensities in the light beam, resulting in an asymmetrical AC-Stark broadened and shifted resonance (Ref. \cite{Sta06}, for example).

\subsection{Comparison with experiment}
The details of the experimental procedure and results of the measurement of the hfs-induced transitions will be presented elsewhere \cite{Eng09}. Briefly, when the two lasers driving the $a\to b$ transition are detuned in frequency from each other (up to 60 GHz in our experiment), we observe a spectral profile (Fig. \ref{Fig:modelfit}, top trace) with peaks evident for all the isotopic (for the isotopes with abundance in excess of 1\%) and hyperfine components of the transition. As the frequencies of the two lasers are tuned towards the same value, the signal decreases in proportion to the inverse square of the frequency detuning of the two lasers. However, at the point of degeneracy, while there is no trace remaining of the zero-spin isotopes, weak lines remain standing for the nonzero-spin isotopes. The intensity of these lines corresponds to a suppression of $\sim 10^9$ compared to an allowed two-photon transition.

The observations are in qualitative agreement with the theoretical analysis presented in this work, assuming $\sR_c\equiv||d_{nc}||/||d_{nb}||\cong1$, $\sR_{c'}\equiv||d_{nc'}||/||d_{nb}||\cong0.7$, and that the mixing $A^{\kappa}\ts{Mix}$ and splitting $A^{\kappa}\ts{Split}$ amplitudes are of opposite sign. In future work, we will perform quantitative analysis of the intensity ratios of various hfs-induced transitions.  These measurements will allow us to measure the off-diagonal hyperfine-mixing parameters, and compare them with the forthcoming atomic-structure calculations.  We note that this technique provides a way to measure the phase of the admixed configuration.

\section{Magnetic-field-induced transitions}

The amplitude of a two-photon transition in the presence of an external magnetic
field $\bm{B}=B_0\hat{\bm{z}}$ is a special case of a three-photon
amplitude, where the third photon corresponds to the static magnetic field. A general consideration of such amplitudes is rather
cumbersome, and the irreducible tensor formalism does not appear particularly useful here. Therefore, we write the two-photon transition amplitude in a
reducible form (Ref.\ \cite{Sob92}, Sec.4.3.6):
 \begin{align}\label{Bfield1}
 W_{b,a}&=\sum_{q_1,q_2} (-1)^{q_1+q_2}A_{q_1,q_2}
 \epsilon^1_{-q_1}\epsilon^2_{-q_2}\,,
 \\ \label{Bfield2}
 A_{q_1,q_2}
 &=\sum_{n}
 \frac{\langle b|d_{q_1}|n\rangle \langle n|d_{q_2}|a\rangle
      +\langle b|d_{q_2}|n\rangle \langle n|d_{q_1}|a\rangle}
 {{\frac{E_b+E_a}{2}-E_{n}}}\,.
 \end{align}
As above, we can restrict the sum of Eq.\ \eqref{Bfield2} to the magnetic
sublevels of the single intermediate state. We now neglect hyperfine
structure, but account for the Zeeman splitting of the intermediate
state: $E_n \rightarrow E_n +\mu_0 g_n B_0 M_n$. Here
$\mu_0$ is Bohr magneton and $g_n$ is the Land\'{e} factor of the state
$n$ (for the dominant intermediate state $6s6p\ ^1P_1$ for our barium case, $g_n=1.02$ \cite{Cur2004}).

Expanding the amplitude of Eq.\ \eqref{Bfield2} up to the linear terms in magnetic field, we get:
 \begin{align} \label{Bfield3}
 A_{q_1,q_2}^{(0)}
 &=
 \frac{||d_{an}||\times ||d_{bn}||}
 {\Delta}
 \sum_{M_n}(-1)^{M_n-M_b}
 \times K,
 \\ \label{Bfield4}
 A_{q_1,q_2}^{(1)}
 &=
 \frac{\mu_0 g_n B_0 \times ||d_{an}||\times ||d_{bn}||}
 {\Delta^2}
 \sum_{M_n}(-1)^{M_n-M_b} M_n
 \times K.
 \end{align}
Here the superscript in parentheses indicate the zeroth and first-order terms, and, as above, $\Delta=(E_b+E_a)/2-E_n$, and we have defined
\begin{equation}
 K = \threej(J_b,-M_b)(1,q_1)(J_n,M_n) \threej(J_n,-M_n)(1,q_2)(J_a,M_a)
 +\threej(J_b,-M_b)(1,q_2)(J_n,M_n) \threej(J_n,-M_n)(1,q_1)(J_a,M_a).
\end{equation}

For the case of $J_a=0$ and $J_b=1$, the sum over $M_n$ for the zero-order
amplitude of Eq.\ \eqref{Bfield3} turns to zero. The first-order
amplitude of Eq.\ \eqref{Bfield4} contains the extra factors $M_n$, and the sum
does not generally vanish for $q_1\ne q_2$. Note that the amplitude of Eq.\ \eqref{Bfield4} is suppressed compared to that of an allowed two-photon transition by a factor on the order of $\mu_0 g_n B_0/\Delta$.

The second mechanism through which a magnetic field induces degenerate two-photon $0\rightarrow 1$ transitions is the mixing of the upper level $b$, $J_b=1$, with levels $c$, $J_c\ne 1$. Since we have assumed that the magnetic field is applied along the quantization axis, only sublevels with the same magnetic quantum number can mix. Moreover, since the magnetic-dipole operator only connects atomic states of the same electronic configuration and term (as is well known, for example, in the context of the selection rules for M1 transitions), the mixing of interest to us only occurs between the components of the upper-state fine structure with different values of the total electronic angular momentum.

Taking into account this mixing, and assuming the upper-state mixing is dominated by just one level $c$, we arrive at the amplitude for the two-photon transition that is first-order in $\mu_0B_0$:
 \begin{align} \label{Bfield5}
 \tilde{A}_{q_1,q_2}^{(1)}
 &=
 \frac{\mu_0 B_0 ||S_{cb}||\times ||d_{an}||\times ||d_{cn}||}
 {\Delta (E_b-E_c)}
 \sum_{M_n}(-1)^{J_c-M_n}\threej(J_c,-M_b)(1,0)(J_b,M_b)
 \\
 \nonumber
 &\times\left[
 \threej(J_c,-M_b)(1,q_1)(J_n,M_n) \threej(J_n,-M_n)(1,q_2)(J_a,M_a)
 +\threej(J_c,-M_b)(1,q_2)(J_n,M_n) \threej(J_n,-M_n)(1,q_1)(J_a,M_a)
 \right]\,.
 \end{align}
Here we have written the magnetic-moment operator as
$\vec{\mu}=-\mu_0(\vec{J}+\vec{S})$ and taken into account that
$\vec{J}$ has only diagonal matrix elements. This leaves us with
the reduced matrix element $||S_{cb}||$ of spin $\vec{S}$. This matrix element is nonzero for the components of the same term
with $\Delta J=1$.

For the Ba transition of present interest, the upper $J_b=1$ state is nominally $5d6d\ ^3D_1$.
The fine-structure ``partner'' state $5d6d\ ^3D_2$  lies only 266 cm$^{-1}$
higher. Assuming pure LS-coupling, we estimate the mixing matrix element:
$||S_{cb}||\approx -3/\sqrt{2}$. There are other closely lying states, however, they belong
to the $6s7d$ configuration and cannot be mixed by magnetic field.

\def\arraystretch{1.4}%
%
%
\begin{table}
\begin{center}
\begin{tabular}{|r@{\,}l|D{.}{.}{1}|D{.}{.}{1}|D{.}{.}{1}|D{.}{.}{2}|D{.}{.}{2}|D{.}{.}{2}|}%
  \hline
  && \multicolumn{3}{c|}{$W_{a,b}\tu{Spl}$}&\multicolumn{3}{c|}{$W\tu{Mix}_{a,b}$} \\\hline
  $\hat{\epsilon}_1$ & $\hat{\epsilon}_2$&\multicolumn{1}{c|}{$m_b=+1$}&\multicolumn{1}{c|}{0}&\multicolumn{1}{c|}{$-1$}&\multicolumn{1}{c|}{$+1$}&\multicolumn{1}{c|}{0}&\multicolumn{1}{c|}{$-1$}\\\hline
$\hat x$&$\hat z$& 2.4 && -2.4 &-0.71&&0.71 \\
$\hat z$&$\hat z$& &&& &-1.34&\\
$\hat{\sigma}_+$&$\hat{\sigma}_-$ & &6.7&& &-0.67&\\
$\hat{\sigma}_+$&$\hat{\sigma}_+$ & &&& &&\\
  \hline
\end{tabular}
\end{center}
\caption{Resonant degenerate two-photon amplitude for $J_a=0\:\to\:J_b=1$ transition in the presence of a 1 kG magnetic field directed along the light-propagation axis.  The units are the same as in Table \ref{Tab:Amps}.  Untabulated amplitudes are zero.}
\label{Tab:ZeeAmp}
\end{table}

In principle, two-photon transitions induced by stray magnetic fields could lead to false systematic signals in the experiments testing Bose-Einstein statistics for photons. However, in the current experiment, the magnetic field is too feeble to create a problem at the present level of sensitivity. On the other hand, applying a stronger magnetic field, the effect can be used to calibrate the apparatus without the need to adjust the lasers. In addition, it provides an additional tool for measuring isotope shifts and hyperfine splittings, as well as for spectral-line identification.

\section{Conclusions and outlook}

In conclusion, we have developed a theory of the hyperfine-interaction-induced two-photon transitions that have recently been observed in experiments \cite{Eng09} searching for small violations of Bose-Einstein quantum statistics for photons. There are two distinct physical mechanisms by which the hyperfine-induced transitions arise, which can be distinguished by measuring the relative intensities of the hyperfine-structure components of the transition. Note that the transition amplitude related to the hyperfine splitting is calculated form the known hyperfine-structure constants of the intermediate state, and can thus be used to calibrate the measurement of the off-diagonal hyperfine mixing in the upper state. We were also able to directly calculate the latter effect for the relevant transition in Ba relating it to the hyperfine-structure splitting in one of the excited states. We propose the use of these transition for measuring off-diagonal hyperfine mixing parameters that could constitute a powerful test of atomic-structure calculations for complex atoms.

Additionally, we have considered the degenerate two-photon transitions which, rather than being induced by hyperfine interactions, are induced by an external magnetic field. Again, there are two mechanisms that lead to such transitions: Zeeman splitting of the intermediate state and off-diagonal mixing in the final state.

\section*{Acknowledgements}
We have benefited from discussions with J.\ J.\ Curry, D.\ P.\ DeMille, and R.\ Marrus. This research has been supported by NSF and by the Foundational Questions Institute (FQXi.org).

\bibliography{HFS-induced2phot}
\end{document}